\documentclass[11pt]{article}

\usepackage[a4paper,margin=1in]{geometry}

\usepackage[T1]{fontenc}
\usepackage[utf8]{inputenc}
\usepackage{lmodern}

\usepackage{xcolor}
\usepackage{microtype}
\usepackage{graphicx}
\usepackage{svg}          
\usepackage{titlesec}
\usepackage{titling}
\usepackage[backref]{hyperref}
\usepackage[numbers,sort&compress]{natbib}
\usepackage[font=small,labelfont=bf]{caption}
\usepackage{fancyhdr}
\usepackage{subcaption}
\usepackage[most]{tcolorbox}

\usepackage{amsmath,amssymb,mathtools,amsfonts}
\usepackage{algorithm}
\usepackage{algorithmic}
\usepackage{subcaption}
\usepackage{booktabs}
\usepackage{nicefrac}

\definecolor{IcebergBlue}{HTML}{6B8DB5}      
\definecolor{IcebergLight}{HTML}{C7C9D1}     
\definecolor{IcebergDark}{HTML}{1B5E98}      
\definecolor{IcebergCyan}{HTML}{89B8C2}      
\definecolor{IcebergBg}{HTML}{F8F9FB}        
\colorlet{AccentColor}{IcebergDark}
\definecolor{IcebergVeryLight}{HTML}{E9F2FA} 

\hypersetup{colorlinks=true,linkcolor=AccentColor,urlcolor=AccentColor,citecolor=AccentColor}

\makeatletter
\renewcommand{\and}{\hspace{1em}}
\makeatother

\titleformat{\section}{\Large\sffamily\bfseries\color{AccentColor}}{\thesection}{1em}{}
\titleformat{\subsection}{\normalsize\sffamily\bfseries\color{IcebergDark}}{\thesubsection}{1em}{}
\titlespacing*{\section}{0pt}{1.2em}{0.6em}
\titlespacing*{\subsection}{0pt}{0.8em}{0.4em}

\DeclareCaptionLabelFormat{accent}{\textcolor{AccentColor}{\bfseries #1~#2}}
\newtcolorbox{callout}{colback=IcebergBg,colframe=AccentColor!80!black,boxrule=0pt,arc=2pt,left=6pt,right=6pt,top=6pt,bottom=6pt}

\newcommand{\accentrule}{\begingroup\color{AccentColor}\hrule height 0.6pt\endgroup}
\newcommand{\titleskip}{0.8em} 

\usepackage{tikz}

\pretitle{%
  \vspace*{-15mm}
  {\raggedright\color{AccentColor}\sffamily\textbf{PROJECT ICEBERG}}
  \vspace{0.2em}
  \accentrule
  \vspace{1em}
  \begin{center}\Huge\sffamily\bfseries\color{black}
}

\posttitle{\par\end{center}}
\preauthor{\vspace{\titleskip}\begin{center}\large\sffamily}
\postauthor{\par\end{center}\vspace{\titleskip}}
\predate{\vspace{0pt}}
\postdate{\vspace{0pt}}

\title{The Iceberg Index: \\ Measuring Skills-centered Exposure in the AI Economy}

\author{Ayush Chopra$^{1,6}$ \and Santanu Bhattacharya$^{1,6}$ \and DeAndrea Salvador$^{3,4,1}$ \and Ayan Paul$^{6}$ \and Teddy Wright$^{5}$ \and Aditi Garg$^{6}$ \and Feroz Ahmad$^{6}$ \and Alice C. Schwarze$^{5}$ \and Ramesh Raskar$^{1,6}$ \and Prasanna Balaprakash$^{2}$ \\
[0.6em] \small $^{1}$Massachusetts Institute of Technology \qquad $^{2}$Oak Ridge National Laboratory \\ [0.2em] \qquad $^{3}$North Carolina General Assembly \qquad $^{4}$Future Caucus: National Task Force on State AI Policy \\ [0.2em] \qquad $^{5}$Utah Office of AI Policy, Dept of Commerce  \qquad $^{6}$Project Iceberg}

\date{\vspace{0pt}}

\begin{document}

\thispagestyle{empty}
\maketitle

\pagestyle{fancy}

\begin{tcolorbox}[
  enhanced,                               
  colback       = IcebergBg,
  colframe      = AccentColor!80!black,
  boxrule       = 0pt,
  arc           = 2pt,
  before skip   = 0pt,          
  after skip    = \titleskip,    
]
\noindent\textbf{Abstract.} 
Artificial Intelligence is reshaping America's over \$9.4 trillion labor market, with cascading effects that extend far beyond visible technology sectors. When AI automates quality control in automotive plants, consequences spread through logistics networks, supply chains, and local service economies. Yet traditional workforce metrics cannot capture these ripple effects: they measure employment outcomes after disruption occurs, not where AI capabilities overlap with human skills before adoption crystallizes. Project Iceberg addresses this gap using Large Population Models to simulate the human–AI labor market, representing 151 million workers as autonomous agents executing over 32,000 skills across 3,000 counties and interacting with thousands of AI tools. It introduces the Iceberg Index, a skills-centered metric that measures the wage value of skills AI systems can perform within each occupation. The Index captures technical exposure, where AI can perform occupational tasks, not displacement outcomes or adoption timelines. Analysis shows that visible AI adoption concentrated in computing and technology (2.2\% of wage value, approximately \$211 billion) represents only the tip of the iceberg. Technical capability extends far below the surface through cognitive automation spanning administrative, financial, and professional services (11.7\%, approximately \$1.2 trillion). This exposure is fivefold larger and geographically distributed across all states rather than confined to coastal hubs. Traditional indicators such as GDP, income, and unemployment explain less than 5\% of this skills-based variation, underscoring why new indices are needed to capture exposure in the AI economy. By simulating how capabilities may spread under alternative scenarios, Project Iceberg enables policymakers and business leaders to identify exposure hotspots, prioritize training and infrastructure investments, and test interventions before committing billions to implementation. Iceberg is built with the AgentTorch framework.
\end{tcolorbox}

\begin{tcolorbox}[colback=white,colframe=IcebergDark]
\centering
\textbf{The Iceberg Index:} A skills-centered KPI for the AI economy. \\
It measures the percentage of wage value of skills that AI systems can perform within each occupation, revealing where human and AI capabilities overlap.
\end{tcolorbox}


\section{Introduction}
Artificial Intelligence is reshaping America’s \$9.4 trillion labor market, creating cascading effects across industries and communities that extend beyond visible technology sectors. When AI systems automate quality control in automotive manufacturing, the consequences propagate through supplier networks, logistics operations, and local service economies. These ripple effects multiply the economic stakes of AI adoption, yet remain invisible to the systems that guide billion-dollar workforce and infrastructure decisions.

While consumer applications such as ChatGPT capture public attention - helping with homework, writing, or quick research - the larger restructuring is already underway inside firms. AI systems now generate more than a billion lines of code each day, prompting companies to restructure hiring pipelines and reduce demand for entry-level programmers~\cite{TrueUp2025}. These observable changes in technology occupations signal a broader reorganization of tasks that extends beyond software development.

AI capabilities now span diverse operational contexts. Financial institutions deploy AI for document processing and analytical support. Healthcare systems automate administrative tasks, enabling clinical staff to allocate more time to patient care. Logistics operations integrate AI-powered systems to optimize fulfillment while creating demand for maintenance and coordination roles. Manufacturing facilities deploy AI-driven quality control to automate inspection tasks while creating roles in system coordination and oversight. Analysis of Bureau of Labor Statistics skill taxonomies reveals that current AI systems can technically perform approximately 16 percent of classified labor tasks (see Appendix A) - yet this technical capability remains largely invisible to the workforce planning systems guiding billion-dollar investments.

Historical precedent demonstrates both the opportunities and risks of such technological shifts. The internet revolution generated substantial economic value: technology companies now represent over one-third of S\&P 500 market capitalization~\cite{ssga_spy_2025} and the U.S. digital economy contributes approximately \$4.9 trillion (\~18\%) to GDP~\cite{bea_2023_infographic}. The impact extends beyond technology firms - research indicates each high-technology job correlates with approximately five additional positions in local services, healthcare, and retail ~\cite{moretti2012,brookings2012}. Early-moving regions captured durable advantages during the internet era: North Carolina’s Research Triangle matured into a global research hub~\cite{rtp_case_2018}; Texas scaled Austin into a top tech market~\cite{cbre_2024_austin}; Tennessee and Kentucky became national logistics leaders as FedEx’s Memphis super-hub and UPS Worldport in Louisville expanded~\cite{memphis_airport_cargo,ups_worldport_2022}; and Utah’s “Silicon Slopes” rose as a cloud computing center~\cite{cbre_2024_slc}.

Similar dynamics will now shape AI adoption patterns. Regions that align skill development, infrastructure investment, and industry strategy early may establish competitive advantages, while delayed response could result in widening disparities. Project Iceberg provides analytical infrastructure to support evidence-based planning as AI capabilities expand across the economy.



\begin{figure}[t!]
    \centering
    \includegraphics[width=0.97\linewidth]{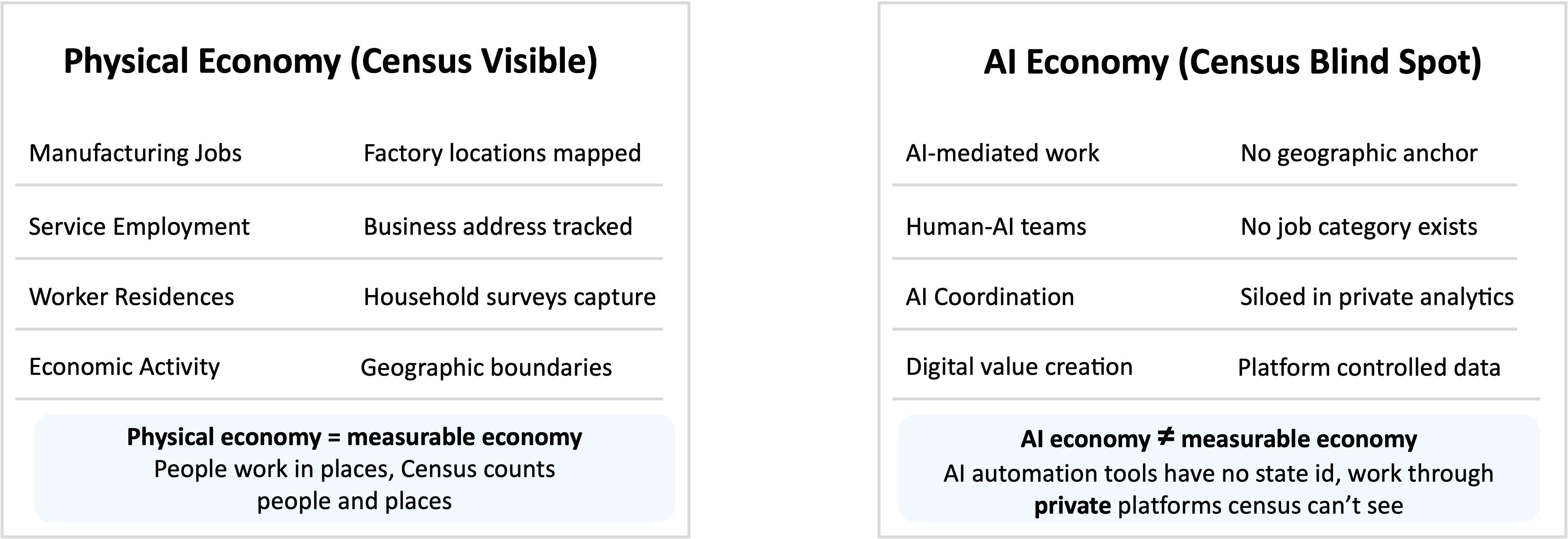}
    \caption{Traditional workforce metrics miss AI-mediated tasks. Census data captures jobs tied to geographic locations and business addresses. Human-AI collaboration—where workers and AI systems jointly perform tasks within occupations—creates new forms of labor that existing metrics don't capture. The Iceberg Index provides a forward-looking measure of this technical automation exposure, revealing skill-based transformation that remains invisible.}
    \label{fig:placeholder}
\end{figure}

\section{Measurement Gaps in Workforce Planning}
Federal and state governments are already committing billions to prepare for the AI economy. The federal AI Action Plan~\cite{whitehouse2025ai} outlines 90 policy positions and has launched a national AI Workforce Research Hub~\cite{whitehouse2025ai}. DOE has identified 16 federal sites for data center development~\cite{doe2025sites}. 

States are responding with major initiatives: 
\begin{itemize}
    \item North Carolina has secured a \$10 billion investment to expand data center capacity~\cite{amazon2025nc}
    \item Tennessee's Google-Kairos reactor will power data centers while anchoring Oak Ridge as a nuclear innovation hub~\cite{google2025tennessee}
    \item Utah's Operation Gigawatt will double statewide clean-energy production over the next decade~\cite{utah_operation_gigawatt_2025}
    \item Virginia has committed \$1.1 billion to train 32,000 AI graduates~\cite{vedp2025datacenters}
    \item DOE has also committed \$100 million for nuclear safety training as that workforce is projected to triple by 2050~\cite{doe2025nuclear}
\end{itemize}
These efforts reflect real momentum in infrastructure, energy, and talent development.

\noindent Yet evidence suggests workforce change is occurring faster than planning cycles can accommodate. Payroll data covering millions of workers shows a 13\% relative decline in early-career employment (ages 22–25) for AI-exposed occupations~\cite{stanford_workforce}. Analysis of job postings across 285,000 firms for 62 million workers reinforces the pattern: the demand for entry-level positions
have subdued while the focus has shifted to hiring for experienced roles~\cite{harvard_workforce}. These shifts, alongside widespread restructuring in the technology sector~\cite{TrueUp2025}, indicate that the pace of change is accelerating across the economy. States are committing billions to workforce programs while key workforce dynamics remain invisible to traditional planning tools.

\subsection{Challenge 1: Anticipating Workforce Shifts Before They Happen}
The labor market is evolving faster than current data systems can capture. AI automates some skills and augments others, creating uneven effects across industries and regions. Much of this activity occurs on digital platforms—gig marketplaces, AI copilots, freelance networks—that fall outside conventional reporting~\cite{oecd_platform_gap_2019}. By the time these changes appear in official statistics, policymakers may already be reacting to yesterday’s disruptions, committing billions to programs that target skills already displaced. Without forward-looking capability to test strategies before implementation, states cannot distinguish investments that prepare workers from those that arrive too late.



\begin{tcolorbox}[colback=white,colframe=IcebergDark]
\centering \textbf{The Invisible Economy:} \\
AI adoption and worker behavior unfold on platforms outside official data. Feedback loops between technology maturity, firm adoption, and state interventions mean that by the time disruptions appear in surveys, policy responses may target skills already displaced or industries already transformed.
\end{tcolorbox}

\subsection{Challenge 2: Measuring Human–AI Work, Not Just Human Work.}
Existing workforce planning frameworks were designed for human-only economies. They track employment, wages, and productivity, but were not designed to measure where AI capabilities overlap with human skills before adoption reshapes occupational structure. Each major transition required new measurement: in the postwar era, output per hour captured physical productivity~\cite{bls_productivity_intro}; in the internet era, the Digital Economy Satellite Account measured the value of online services~\cite{bea_desa_2023}. The AI era is defined by intelligence itself becoming a shared input between humans and machines. What matters is not only the number and complexity skills to execute, but how much of each skill's value can be performed by AI and which human skills remain differentially valuable. Without a skills-centered metric, states lack a systematic way to align investments with where AI-human skill overlap is concentrated.

\begin{tcolorbox}[colback=white,colframe=IcebergDark]
\centering \textbf{Every Revolution Needs a New Metric:} \\
Industrial era $\rightarrow$ Output per hour (measured physical productivity) \\
Internet era $\rightarrow$ Digital economy accounts (captured online service value) \\
Intelligence era $\rightarrow$ Skills-centered measure (reveals AI-human skill overlap)
\end{tcolorbox}

\section{Project Iceberg: Sandbox for the Human-AI Workforce}
Project Iceberg simulates the emerging human-AI workforce. It models how 151 million American workers and emerging AI capabilities interact, allowing states to explore policy scenarios and assess potential workforce exposure patterns before committing billions to infrastructure and training programs. Built on MIT’s Large Population Models~\cite{chopra2025lpm} and powered by Oak Ridge National Laboratory’s Frontier supercomputer~\cite{dash2023optimizingdistributedtrainingfrontier}, Iceberg turns trillions of workforce data points into scenario-planning capability.

\begin{figure}[t!]
    \centering
    \includegraphics[width=0.99\linewidth]{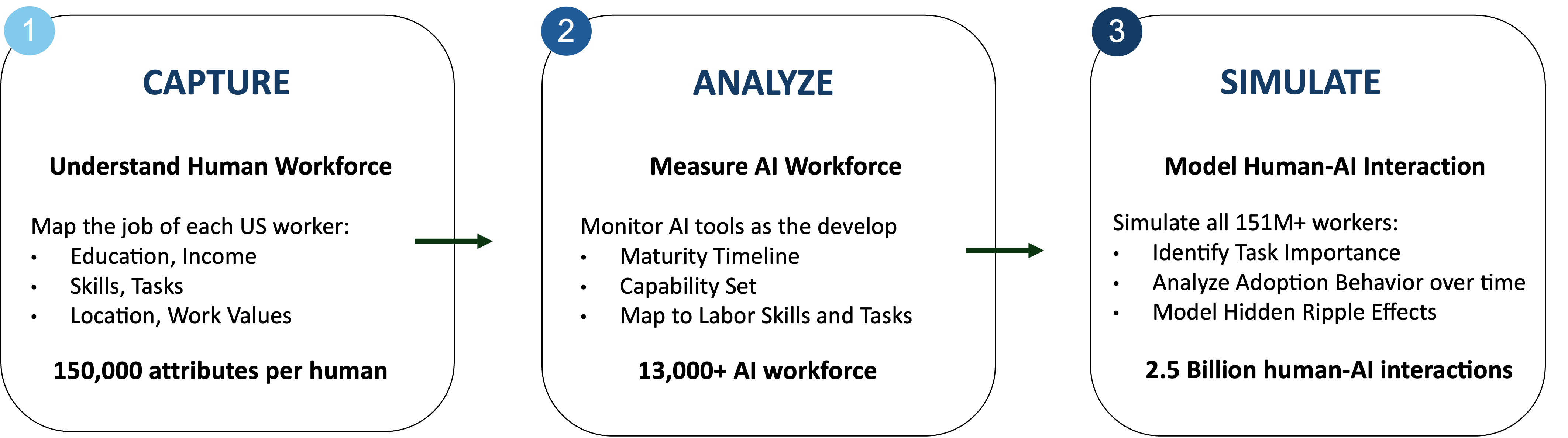}
    \caption{\textbf{Project Iceberg prepares states for the AI economy.} First platform to simulate human-AI workforce interactions at national scale, enabling policymakers to assess technical exposure, test workforce strategies, and target investments before committing billions to implementation.}
    \label{fig:iceberg_process}
\end{figure}

\begin{figure}[t!]
    \centering
    \includegraphics[width=0.99\linewidth]{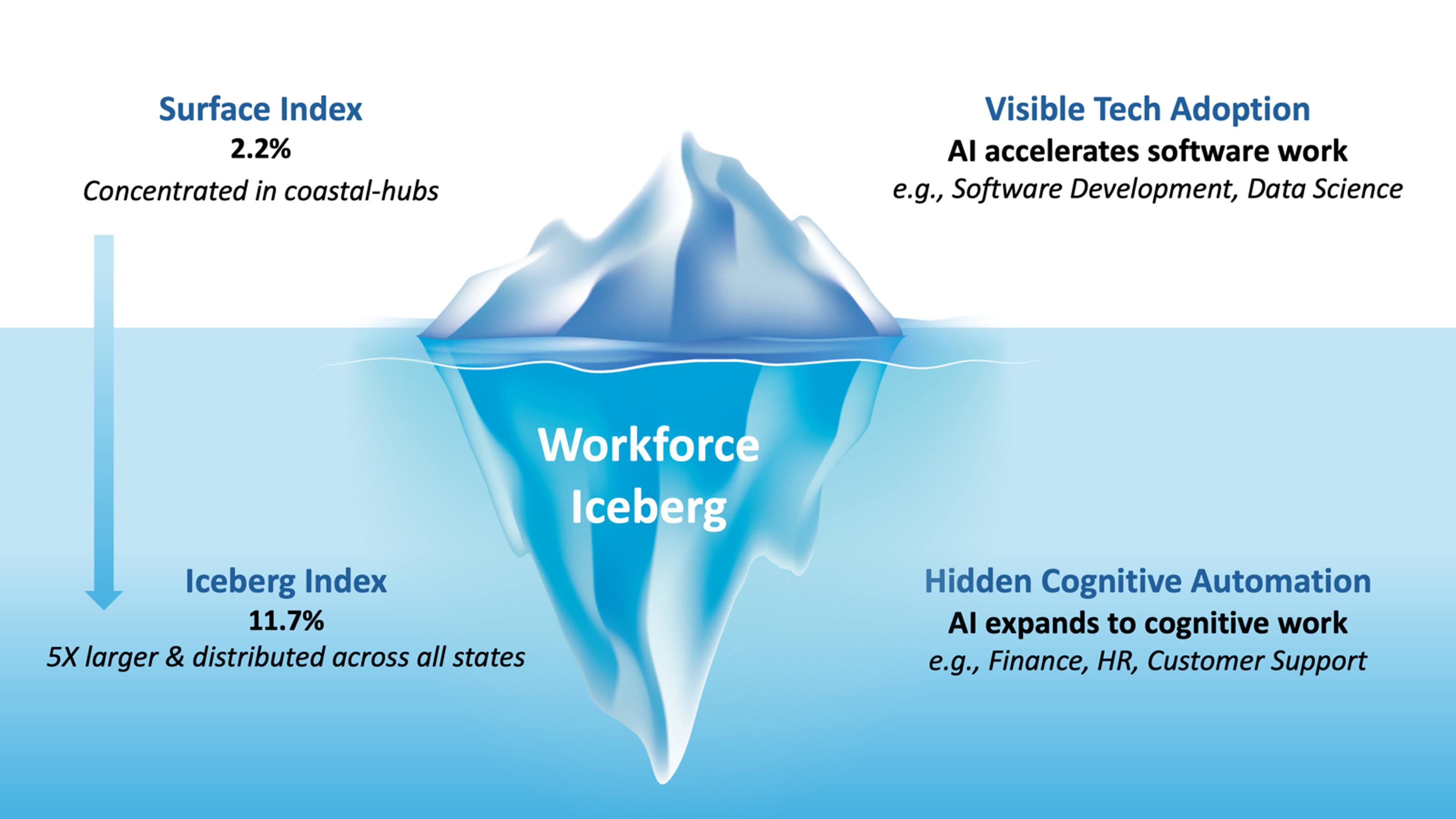}
    \caption{The Iceberg Index reveals workforce exposure five times larger than visible tech adoption. Above the waterline: the Surface Index (2.2\%) tracks current AI adoption in coastal technology hubs. Below the waterline: the Iceberg Index (11.7\%) measures technical capability spanning administrative, financial, and professional services nationwide. This hidden mass represents where workforce preparation strategies based solely on visible tech-sector signals may fall short.}
    \label{fig:iceberg_overview_us}
\end{figure}

Iceberg represents both sides of the emerging labor market: what human workers currently do, and what AI systems are becoming capable of doing. By mapping these onto a shared skill taxonomy, it simulates how work evolves under different assumptions about policy intervention, adoption behavior, and technology readiness. The process unfolds in three steps (Fig~\ref{fig:iceberg_process}):

\textbf{\textcolor{IcebergDark}{Step 1 - Human Workforce Mapping:}} We construct a digital representation of the workforce: 151 million workers across 923 occupations in 3,000 counties, covering more than 32,000 distinct skills~\cite{onet2025,acs2025}. Each worker is modeled as an agent with attributes such as skills, tasks, and location. This enables analysis of reskilling potential and occupational similarity, which is essential for designing pathways when skills and tasks shift.

\textbf{\textcolor{IcebergDark}{Step 2 – AI Workforce Mapping:}} We catalog over 13,000 AI tools - including copilots and workflow automation systems - using the same skill taxonomy. This allows direct comparison between human and AI capabilities. The alignment reveals where AI augments human work (e.g. automating hospital paperwork so nurses can spend more time with patients) and where it transforms tasks entirely (e.g. accelerating code generation requires engineers to shift towards oversight, testing, and integration skills).

\textbf{\textcolor{IcebergDark}{Step 3 - Human--AI Simulation:}} We combine these two populations in Large Population Models (LPMs), simulating billions of interactions between workers, skills, and AI tools. These simulations incorporate factors such as technology readiness, adoption behavior, and regional variation. Policymakers can use the resulting scenarios to anticipate disruption, test reskilling strategies, and allocate resources where they will have the most impact.

\begin{tcolorbox}[colback=white,colframe=IcebergDark,title={Iceberg Simulations: Scale and Infrastructure}]
\begin{itemize}
    \item \textbf{Workforce Scale:} 151 million workers across 923 occupations in 3,000 counties
    \item \textbf{Skill Coverage:} 32,000+ distinct skills from standardized taxonomies
    \item \textbf{AI Systems:} 13,000+ tools across software, cognitive, and physical capabilities
    \item \textbf{Simulation Engine:} Large Population Models used for national security~\cite{chopra2025lpm}.
    \item \textbf{Computing Platform:} Frontier supercomputer at Oak Ridge National Lab~\cite{frontier2022}.
\end{itemize}
\end{tcolorbox}

Once constructed, leaders can adjust inputs—such as training programs, incentive structures, or regulatory choices—and run scenarios under different assumptions of technology maturity and adoption. The simulation then yields three kinds of insight: how occupations and skills evolve over time, where disruption is geographically spread, and how shocks in one sector cascade into others. These outputs allow policymakers to compare strategies side by side and anticipate both direct and indirect consequences.

The framework builds on validated foundations (see Appendix A), ensuring that the simulated dynamics reflect real labor-market structure. Its primary output is the Iceberg Index—a skills-centered measure of workforce transformation introduced in Section~\ref{sec:iceberg_index}.

\begin{tcolorbox}[colback=white,colframe=IcebergDark]
\centering \textbf{Project Iceberg Goal:} \
To provide a sandbox for testing workforce strategies before implementation—enabling policymakers to assess technical exposure, evaluate interventions under different scenarios, and target preparation where it matters most.
\end{tcolorbox}

\section{The Iceberg Index}
\label{sec:iceberg_index}
The Iceberg Index is a skills-centered measure of workforce exposure in the AI economy. It quantifies the wage value of occupational tasks that AI systems can technically perform, revealing where human work and AI capabilities overlap. This provides forward-looking intelligence to complement traditional workforce metrics, which measure employment outcomes after disruption occurs rather than technical capability before adoption crystallizes.

\begin{tcolorbox}[colback=white,colframe=IcebergDark]
\centering
\textbf{The Iceberg Index:} A skills-centered KPI for the AI economy. \\
It measures the wage value of skills that AI systems can perform within each occupation, revealing where human and AI capabilities overlap.
\end{tcolorbox}

\subsection{How the Index Works}
The Index evaluates each occupation along three dimensions: the skills required, the automatability of those skills, and the value of the work in wages and employment. Together these factors yield a consistent measure of technical exposure that can be aggregated across occupations, industries, and regions. Formally, the Index for a given occupation weights each skill by its relative importance, automatability score, and prevalence, producing a single exposure value between 0 and 100\%.

The Index captures the share of tasks that are \textit{technically automatable} based on demonstrated AI capabilities. A skill is considered automatable if tools exist that language models can use to perform relevant tasks. This approach captures automation potential through tool availability and language model tool-use capability. Because the Index is skill-based rather than job-based, it can be consistently compared across geographies, industries, and occupation clusters, making it useful for both national and local decision-making.


\subsection{What the Index Measures}
The Index measures \textit{technical exposure}—where AI capabilities and human skills overlap—not displacement outcomes. For example, financial analysts will not disappear, but AI systems may demonstrate capability across significant portions of document-processing and routine analysis work. This reshapes how roles are structured and which skills remain in demand, without necessarily reducing headcount.

\textbf{Systemic exposure, not task capability}. The Index measures how AI tool availability transforms workforce skills at scale, across million of workers. This differs from capability benchmarks like GDPval~\cite{openai2025gdpval} and APEX~\cite{vidgen2025apex}, which test base language model performance on isolated professional tasks. Both perspectives inform workforce planning: benchmarks establish what models can do, while the Index measures how tool ecosystems enable automation in practice. Future work can incorporate capability benchmark data to strengthen assessments of language model performance.

The Index does not predict job losses, adoption timelines, or net employment effects. Actual workforce impacts depend on firm adoption strategies, worker adaptation, regulatory choices, societal acceptance and broader economic conditions. Policymakers should interpret the Index as a capability map—similar to how earthquake risk zones identify exposure without predicting when events occur—that enables scenario testing and proactive planning.

\subsection{Two Ways to Use the Index}
The Iceberg Index serves two complementary purposes. This paper focuses on establishing the baseline Index, while simulation results exploring adoption dynamics will be presented in future work.

\noindent \textbf{Baseline assessment.} The Index measures maximum technical exposure—the share of occupational skills where AI has demonstrated capability in at least one context. This establishes problem scope without making adoption assumptions. It identifies where skill overlap is concentrated, enables geographic and industry comparisons, reveals structural vulnerabilities in workforce composition, and provides a metric for tracking how capability overlap evolves over time. \textit{Key question: Where might AI capabilities reach if adoption follows patterns observed in technology sectors?}

\noindent \textbf{Simulation input.} The baseline Index feeds into scenario modeling that explores how technical capability might translate to workforce impact. Future work will test adoption assumptions (speed, sector patterns), model skill transferability across occupational contexts, evaluate policy interventions, and compare alternative futures under different technology maturity assumptions. Simulations explore: \textit{Key question: Under specific adoption scenarios, how might baseline exposure materialize into workforce change?}

\subsection{Applications for Policymakers}
The Index enables three complementary strategic functions:

\begin{itemize}
    \item \textbf{Status Quo Assessment:} Establish a baseline of automation potential if no action is taken, helping governments identify where adaptation needs are most concentrated. This makes the risks of inaction visible across industries and regions.
    
    \item \textbf{Strategic Opportunity Identification:} Highlight where AI adoption can advance broader goals while strengthening, rather than displacing, the workforce. For example, automating healthcare administration can free nurses for direct patient care, easing shortages while improving quality.
    
    \item \textbf{Recalibrating Investments:} Update traditional job-multiplier assumptions (e.g., one construction job creating 1.8 allied jobs) for the AI economy, showing how ripple effects change when support functions are automated or augmented. This enables more accurate projections of returns on training, infrastructure, and incentive programs.
\end{itemize}

The Index provides a measurable framework for evaluating policy choices, reframing AI as a planning tool for strategic workforce investments. This enables states to prepare for workforce change before it appears in unemployment data or GDP figures. For detailed validation methodology, see Section~\ref{sec:validation} and Appendix A.

\section{Interpreting the Index}
\label{sec:results}
In the previous section, we defined the Iceberg Index and described how it is constructed. Here we demonstrate how it can be applied to the U.S. workforce. The results quantify technical exposure—the share of occupational tasks that current AI systems can perform as mapped across 151 million workers, 923 occupations, and 3,000+ counties. In this analysis we focus on digital AI tools that can perform technical and cognitive tasks, where technology maturity and deployment patterns are observable. Physical automation through robotics is excluded here but will become increasingly relevant as capabilities mature.

Each subsection illustrates a different way the Index can be interpreted: identifying the visible disruptions that are already occurring, uncovering the larger hidden exposures beneath the surface, detecting blind spots where states risk being caught off guard, and distinguishing between concentrated versus distributed industry patterns. These examples are not forecasts, but demonstrations of how the Iceberg Index can function as a policy laboratory: a tool for exploring alternative futures and guiding investment choices.

\subsection{Validation with Real-world Data}
\label{sec:validation}
We validate our methodology through two tests: first, whether skill-based occupational representations capture genuine labor market structure; second, whether exposure predictions align with actual AI adoption.

\begin{figure}[h!]
    \centering
    \includegraphics[width=0.75\linewidth]{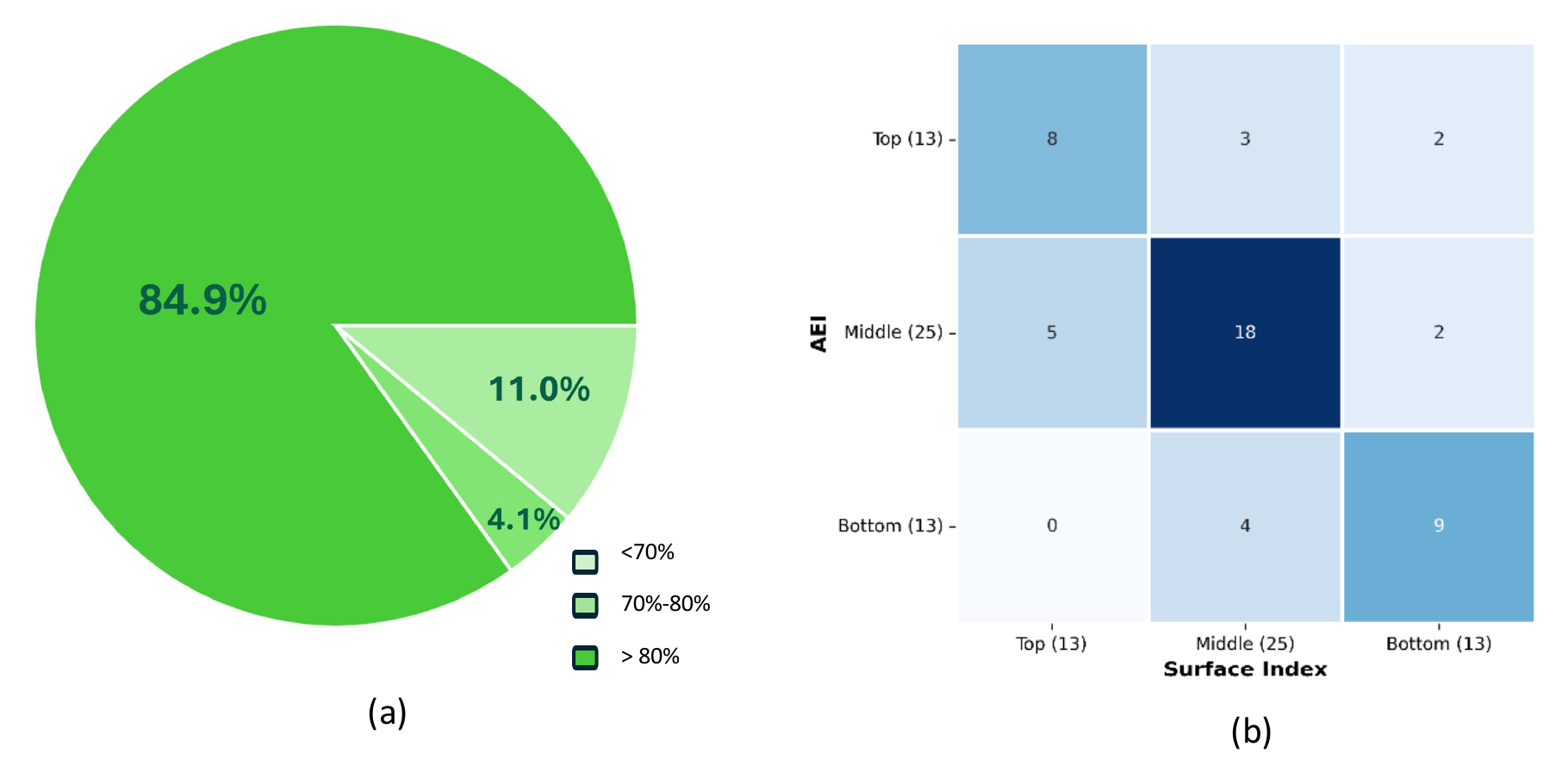}
    \caption{Validation of skill-based framework against empirical data. (a) Skill similarity validation: Pie chart shows prediction accuracy when matching occupation pairs by skill similarity to actual career transitions from O*NET data. For pairs our framework identifies as highly similar (>80\% similarity score, dark green), 85\% correspond to observed career transitions, confirming skill embeddings capture genuine labor market structure. Lower similarity thresholds (70-80\%; <70\%) show decreasing alignment with actual transitions. (b) Adoption validation: State-level comparison shows 69\% agreement between Anthropic Economic Index (actual AI usage patterns) and Surface Index (predicted technical exposure). Strong alignment at both extremes: 8 of 13 leading states and 9 of 13 aspiring states match perfectly, confirming the Index identifies structural exposure before widespread adoption.}
    \label{fig:validation_recall_aei}
\end{figure}

\noindent \textbf{Skill-Based Validation:} We test whether occupations our framework identifies as similar based on skill profiles exhibit similar characteristics in empirical workforce data. Using skill-based embeddings derived from O*NET occupational profiles, we calculate similarity scores between all occupation pairs and compare these against career transition networks that capture observed worker mobility patterns~\cite{onet2024}. Career transition data reflects which occupations workers commonly move between, providing ground truth for occupational relationships independent of our framework. We identify the top percentile of similar occupation pairs according to our skill embeddings and test whether these correspond to frequent career transitions in the empirical data. Our embeddings achieve 85\% recall in predicting these transition relationships (Figure~\ref{fig:validation_recall_aei}(a)): 85\% of commonly observed career moves involve occupations our framework identifies as highly similar based on skills. This high recall rate confirms that skill-based representations capture genuine labor market structure rather than theoretical constructs, validating our approach to measuring AI-human skill overlap using the same framework. \\

\noindent \textbf{Adoption Validation}: Building on this validated skill framework, we test whether automation exposure predictions align with actual AI usage.  The Anthropic Economic Index (AEI) measures real-world AI usage from millions of Claude users nationwide, with industry analysis showing concentration in computing and technical occupations~\cite{appelmccrorytamkin2025geoapi}. We define a Surface Index that models technical exposure in computing and technology tasks across all 50 states. We rank states by both AEI usage and Surface Index exposure. AEI groups states into four adoption tiers; we consolidate the two middle tiers into a single category, creating three balanced groups (leading, emerging, aspiring) to reduce sensitivity to boundary effects. Agreement is measured by whether states fall into the same category in both rankings. We find 69\% geographic agreement, with strong consensus at extremes: 8 of 13 leading states and 9 of 13 aspiring states match perfectly. For instance, Washington, California, and Colorado consistently appear as leaders in both measures, while Wyoming, Mississippi, and Alaska align as laggards. Discrepancies reveal an interpretable pattern: our Index occasionally (18\%) shows higher exposure than current usage in states like Texas and North Carolina, reflecting workforce structures with high technical vulnerability regardless of current adoption levels. However, the framework rarely underestimates—states (13\%) where high AEI usage shows low Index values. This asymmetry validates the Index as a leading indicator: it identifies structural exposure before widespread adoption occurs.

\begin{tcolorbox}[colback=IcebergVeryLight,colframe=IcebergDark,title=\textbf{Iceberg Insight 1: Iceberg Validated with Real-World Data}]
Validation against independent data from millions of AI usage interactions shows strong agreement on leading and aspiring states, confirming our approach captures genuine adoption behavior. Skill-based representations predict 85\% of career transitions, and exposure predictions achieve 69\% geographic agreement with actual usage patterns.
\end{tcolorbox}

\subsection{Quantifying the Tip of the Iceberg}
Building on our validated methodology, we now quantify technical exposure within occupations where AI adoption is currently concentrated. In 2025, more than 100,000 job losses were linked to AI restructuring. AI systems now write over a billion lines of code each day, exceeding human developer output. We measure skill overlap within computing and technology occupations—the Surface Index—which aligns with real-world adoption patterns from millions of AI users (Section~\ref{sec:validation}).

\begin{figure}[h!]
    \centering
    \includegraphics[width=0.85\linewidth]{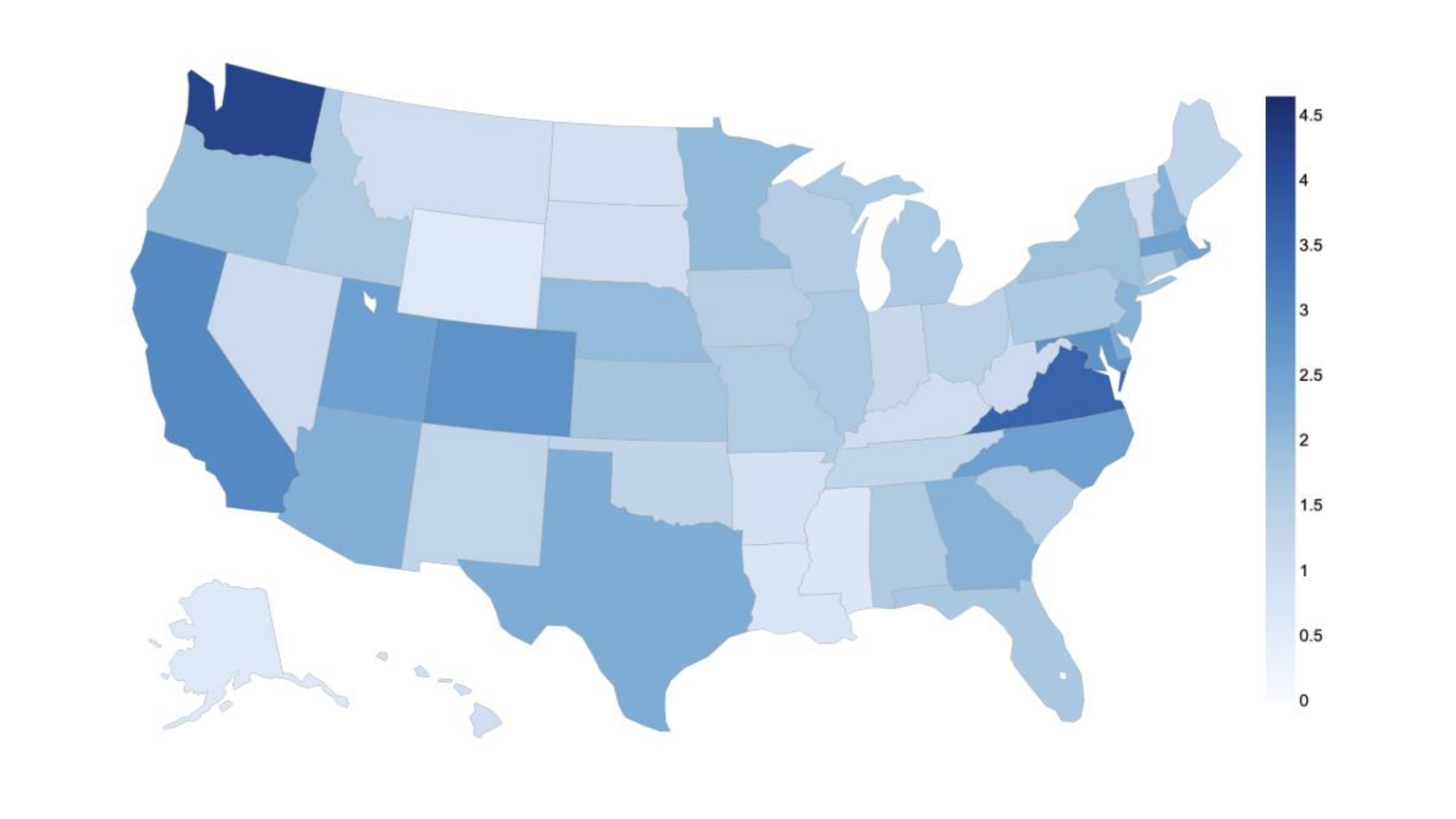}
    \caption{Surface Index validation: Current technology-sector exposure across US states. The average surface index of the country is 2.2, maximum is 4.2 (Washington) and minimum is 0.2 (Guam). 18 states (or territories) are above the national average and 36 are below the national average.}
    \label{fig:surface_index_heatmap}
\end{figure}

Nationally, the Surface Index stands at 2.2\%, representing approximately \$211 billion in wage value across 1.9 million workers in technology occupations (Figure~\ref{fig:surface_index_heatmap}). This includes software engineers, data scientists, analysts, program managers, and related roles where current AI adoption is concentrated. Washington (4.2\%), Virginia (3.6\%), and California (3.0\%) lead in exposure values, yet even in these states, direct technology tasks account for only a small share of employment. The technology sector represents more than 30\% of the S\&P 500's market capitalization but only around 6\% of the workforce, showing that states dominating technology maintain highly diversified labor markets.

States like Mississippi and Wyoming show minimal values on this Surface Index, reflecting their limited technology sector presence and fewer occupations with skill profiles matching current AI adoption patterns. However, this represents only the visible beginning. The full Iceberg Index reveals that these same states show substantial technical exposure in administrative, financial, and professional services—indicating broader potential impact beyond current visible adoption.

\begin{tcolorbox}[colback=IcebergVeryLight,colframe=IcebergDark,title=\textbf{Iceberg Insight 2: Tech Disruption Is Visible but Small}]
Headlines focus on tech skills, but these span occupations representing only 2\% of labor market wage value. The hidden mass beyond visible tech sectors is five times larger.
\end{tcolorbox}

\subsection{The Hidden Mass Beneath the Surface}
Beyond technology occupations, AI capabilities extend to cognitive and administrative work. Tools developed for coding demonstrate technical capability in document processing, financial analysis, and routine administrative tasks - illustrating how capabilities demonstrated in technology contexts translate to other domains. Some adoption is already occurring:  IBM reduced HR staff through AI automation~\cite{ibmnews}, Salesforce froze hiring for non-technical roles~\cite{salesforce}, and McKinsey projects that 30\% of financial tasks could be automated by 2030~\cite{mckinsey2030}.

We apply the same skill-overlap methodology to administrative, financial, and professional service occupations beyond the technology sector. The Iceberg Index for digital AI shows values averaging 11.7\%—five times larger than the 2.2\% Surface Index. Unlike technology-sector exposure concentrated in coastal hubs, this broader skill overlap is geographically distributed. South Dakota, North Carolina, and Utah show higher Index values than California or Virginia.

\begin{figure}[h!]
    \centering
    \includegraphics[width=0.95\linewidth]{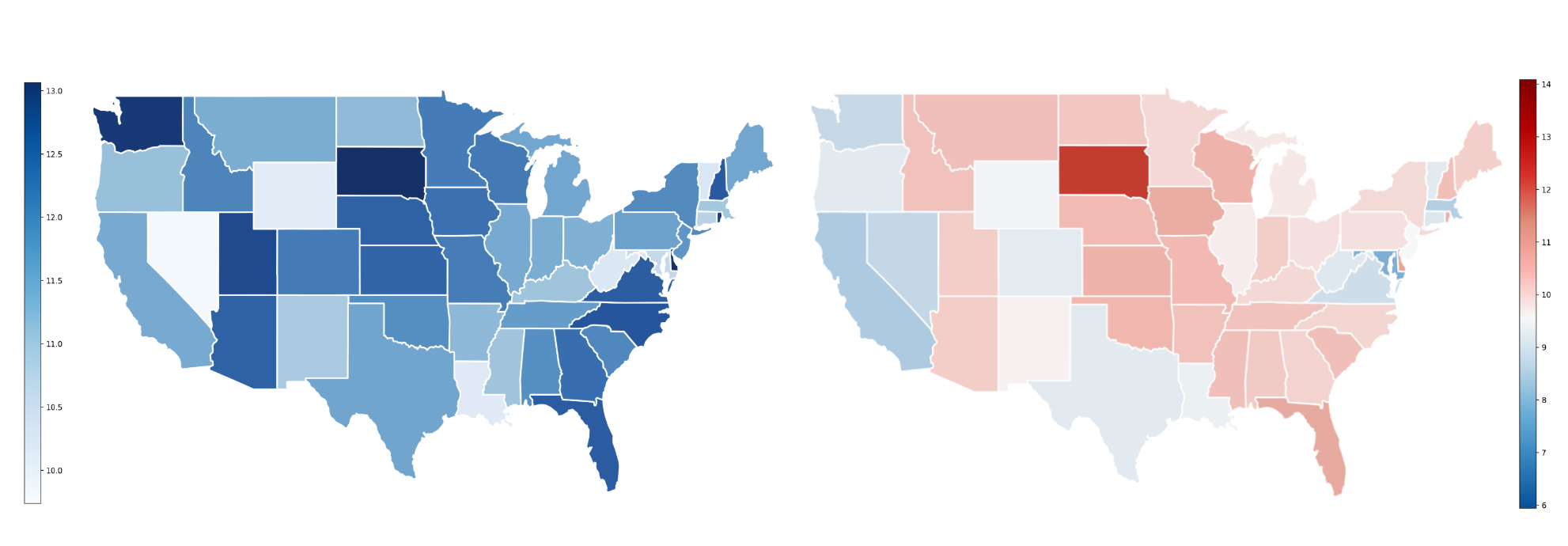}
    \caption{Geographic patterns of hidden automation exposure. (Left) Digital Iceberg Index shows cognitive automation spreads beyond coastal tech hubs, with unexpected leaders like Delaware and South Dakota showing higher exposure than California due to concentrated administrative and financial sectors. (Right) Automation surprise reveals which states face the largest gaps between current visibility and projected transformation. Manufacturing states like Ohio and Michigan show substantial hidden white-collar exposure years before anticipated physical automation, requiring proactive preparation for administrative and coordination role changes.}
    \label{fig:find2_digitalindex}
\end{figure}

Industrial states illustrate this pattern. Tennessee (11.6\%) and Ohio (11.8\%) show substantial Index values driven by administrative and coordination roles within factories and supply chains.
These white-collar functions show technical exposure that maybe invisible to policymakers while states focus largely on physical automation. These patterns reveal where skill overlap extends beyond current visible adoption, though actual workforce impacts will depend on adoption decisions, quality thresholds, and organizational constraints (Figure~\ref{fig:find2_digitalindex}(a)).

\begin{tcolorbox}[colback=IcebergVeryLight,colframe=IcebergDark,title=\textbf{Iceberg Insight 3: White-Collar Exposure Is Nationwide}]
Administrative and financial tasks where AI demonstrates capability span five times more wage value than visible tech disruption—and are geographically distributed across all states, not just coastal. States like Delaware and South Dakota show higher Index values than California.
\end{tcolorbox}

\subsection{The Blind Spot: Automation Surprise}
Some states show large differences between Surface Index (technology occupations) and Iceberg Index values (including administrative, financial, and professional service occupations). These gaps indicate substantial skill overlap in cognitive work despite minimal visible technology-sector adoption, potentially creating mismatches between preparation strategies and technical capability patterns.

America's industrial heartland shows the largest gaps. Rust Belt states such as Ohio, Michigan, and Tennessee register modest Surface Index values but substantial Iceberg Index values driven by cognitive work—financial analysis, administrative coordination, and professional services—that supports manufacturing operations. Tennessee illustrates this pattern: Surface Index of 1.3\% but Iceberg Index of 11.6\% indicating that administrative and service functions show up to ten times greater technical exposure than visible technology occupations. South Dakota follows a similar pattern, combining low technology-sector Index values with high administrative-sector overlap. Figure~\ref{fig:find2_digitalindex}(b) maps these gaps across states.

By contrast, technology-intensive states such as California and Washington show high values on both Surface and Iceberg Indices, resulting in smaller gaps. These states may recognize preparation needs earlier because skill overlap is already visible in their dominant technology sectors. The gaps emerge because current workforce planning often focuses on visible technology-sector adoption, while cognitive and administrative work has received less attention in preparation strategies. As a result, states with small technology sectors underestimate the scale of their exposure, leaving them vulnerable when adoption accelerates in white-collar work.



\begin{tcolorbox}[colback=IcebergVeryLight,colframe=IcebergDark,title=\textbf{Iceberg Insight 4: Manufacturing States Face Hidden White-Collar Surprise}]
Midwest states like Ohio and Michigan face double-digit technical exposure in white-collar work while states may focus on physical automation. Cognitive and administrative exposure from validated AI capabilities measures upto ten times higher than technology-sector exposure.
\end{tcolorbox}

\subsection{Industry Patterns: Concentrated vs. Distributed Impact}
States with similar Iceberg Index values face very different challenges depending on how that exposure is distributed across industries. In some cases, exposure is tightly concentrated in just a few dominant sectors. In others, the same level of exposure is spread broadly across many parts of the economy. This structure matters because it determines how states must respond: distributed risk demands broad, multi-sector coordination, while concentrated risk allows for targeted, sector-specific action.

To distinguish between these patterns, we calculate each industry's contribution to a state's total Iceberg Index value and compute the Herfindahl-Hirschman Index (HHI) of this distribution. The HHI quantifies whether Index values concentrate in a few industries or spread broadly across sectors and is a standard measure adapted from labor economics~\cite{Azar2020LaborMarketConcentration, RevisitedHHI2023}. We rank all states by their HHI values and divide them into thirds based on the observed distribution: the bottom third (HHI $\leq$ 1580) represents the most distributed patterns, the middle third (1581–1737) shows moderate concentration, and the top third ($\geq$ 1738) represents the most concentrated patterns. These categories are descriptive classifications based on relative position in the data, not normative thresholds. Details appear in Appendix B and Figure~\ref{fig:industry_geoplot} visualizes these patterns geographically.

\begin{figure}[h!]
    \centering
    \includegraphics[width=0.67\linewidth]{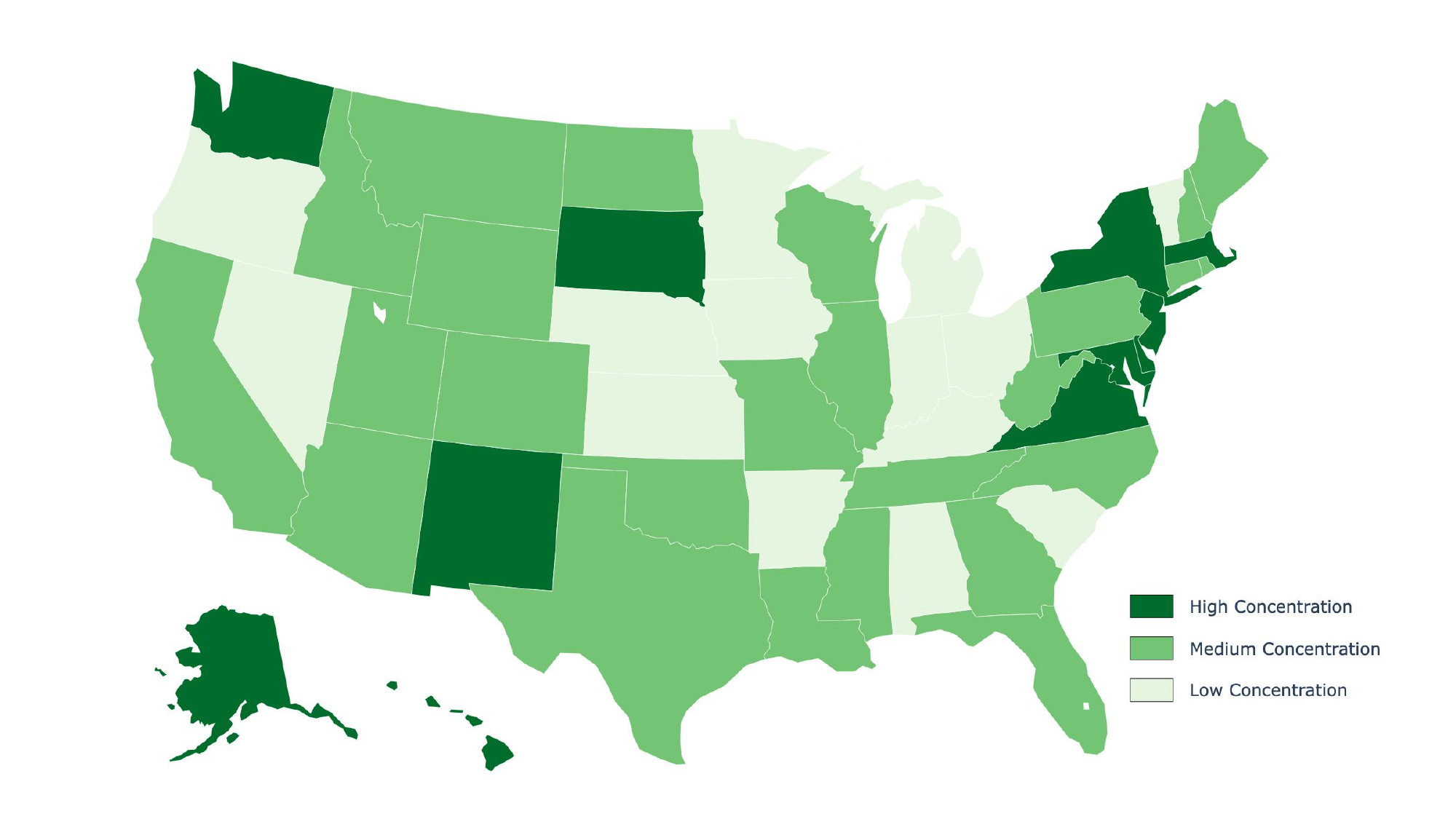}
    \caption{Geographic distribution of automation exposure concentration. States are grouped into three tiers based on their Herfindahl–Hirschman Index (HHI) computed from how Iceberg Index values distribute across industries within each state. The national HHI range spans from 1422 (most distributed) to 1895 (most concentrated). We divide states into thirds by HHI ranking: Most Distributed ($\leq$ 1580, light green), Moderate (1581–1737, medium green), and Most Concentrated ($\geq$ 1738, dark green). This classification reveals regional contrasts, with distributed patterns across the Manufacturing Belt and concentrated patterns along the Northeast corridor.}
    \label{fig:industry_geoplot}
\end{figure}

The analysis reveals clear regional contrasts. Northeastern belt tend to show concentrated exposure, especially in finance and technology. In contrast, states across the Manufacturing Belt display more distributed patterns, with exposure spread across logistics, production, administration, and services. These structural differences influence how disruption might propagate — whether driven by a few sectors or ripple broadly across the economy.

Even when overall Index values appear similar, underlying structures can vary significantly. For example, Iowa (12.22\%) and Ohio (11.34\%) both show broadly distributed patterns, while Virginia (12.48\%) channels a comparable level of exposure dominated by just two sectors: finance and technology. The same Iceberg Index can imply very different workforce vulnerabilities depending on how it is composed.

\begin{tcolorbox}[colback=IcebergVeryLight,colframe=IcebergDark,title=\textbf{Iceberg Insight 5: Structure of Exposure Determines Strategy}]
The same level of technical exposure can require entirely different responses. Concentrated patterns enables sector-specific action, while distributed patterns demands multi-sector coordination
\end{tcolorbox}

\subsection{Why Traditional Metrics Miss the Iceberg}
\label{sec:why_iceberg}
Traditional economic metrics—GDP, per-capita income, and unemployment—are widely used to benchmark state performance. We test how these measures align with workforce exposure as captured by the Iceberg Index. Using Q2 2025 data, we rank all 50 states by GDP, income, and unemployment, and then compare these with state rankings from the Index: first for the Surface Index (today’s visible disruptions concentrated in technology and software tasks) and then for the Iceberg Index (a forward-looking measure of the projected spread to cognitive and administrative tasks).

\begin{figure}[t!]
    \centering
    \includegraphics[width=0.70\linewidth]{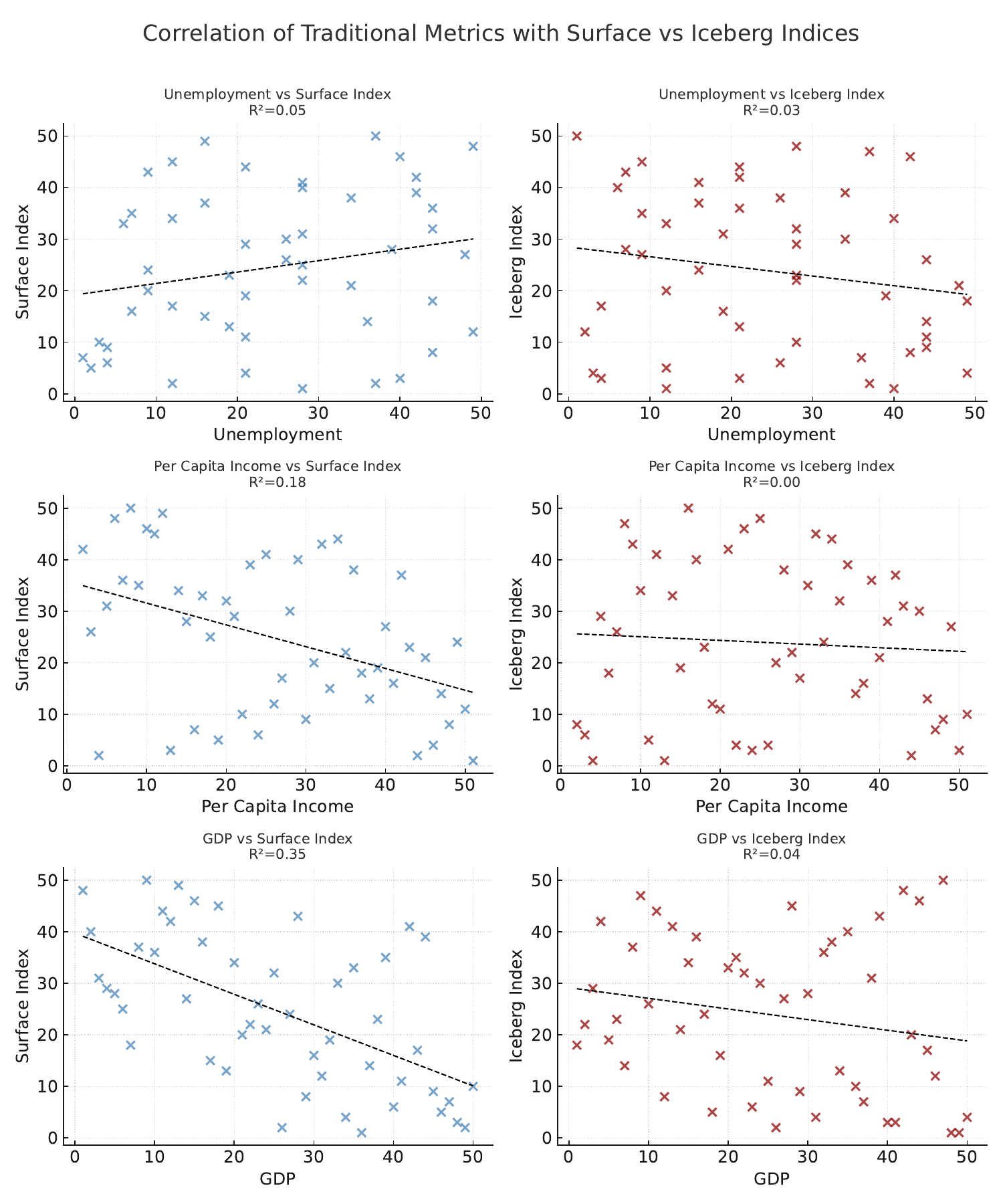}
    \caption{Correlations of traditional economic metrics with the Iceberg Index. Each panel shows the relationship between a conventional signal (unemployment, per-capita income, or GDP) and state-level automation exposure. The top row reports the \textbf{Surface Index}, reflecting today’s visible adoption concentrated in technology. The bottom row reports the \textbf{Iceberg Index}, projecting systemic exposure across finance, administration, and professional services. Regression lines and $R^{2}$ values quantify explanatory power. Traditional metrics correlate with the Surface Index but fail to explain the Iceberg Index, highlighting why policymakers relying on unemployment 
    or GDP will misjudge hidden exposure.}
    \label{fig:correlation_scatterplots}
\end{figure}

The results show clear differences. Traditional metrics exhibit moderate alignment with the Surface Index: states with higher GDP or lower unemployment often have larger technology sectors and therefore higher visible exposure today. By contrast, their relationship with the Iceberg Index is negligible, with GDP, income, and unemployment each explaining less than five percent of the variation in systemic  exposure; in some cases, the correlations are weakly negative (Figure~\ref{fig:correlation_scatterplots}). For example, Delaware and Utah exhibit higher Iceberg exposure than California, despite much smaller economies, because their concentrated finance and administrative sectors present sharper automation targets than California’s diversified workforce. This reflects a broader pattern: technology automation is clustered in coastal hubs, while cognitive automation spans administrative, financial, and professional roles present in every state.

These findings underscore the need for a complementary benchmark. GDP, income, and unemployment remain essential for assessing economic performance, but they were not designed to capture how skills and tasks itself is changing in the human–AI economy. The Iceberg Index fills that gap by providing a forward-looking KPI that measures skills-centered exposure directly. By using both traditional metrics and the Iceberg Index, policymakers gain a more complete picture of today’s economy and tomorrow’s transitions, enabling better alignment of training, infrastructure, and investment strategies with the realities of AI adoption.

\begin{tcolorbox}[colback=IcebergVeryLight,colframe=IcebergDark,title=\textbf{Iceberg Insight 6: Traditional Metrics Miss the Hidden Risk}]
GDP and unemployment track today’s visible tech disruption, but they fail to capture the nationwide spread of white-collar automation revealed by the Iceberg Index.
\end{tcolorbox}
\section{Limitations}
The Iceberg Index captures technical exposure—where AI capabilities overlap with occupational skills—not actual displacement or labor market outcomes. Real-world impacts depend on adoption choices, firm strategies, worker adaptation, societal acceptance and policy interventions. The Index functions as a capability map enabling scenario-based planning rather than deterministic forecasting.

\subsection{Validation}
The framework achieves strong empirical validation against independent data sources: high recall rates predicting career transitions and substantial geographic agreement with observed AI adoption patterns (see Section~\ref{sec:validation} for detailed results). These results confirm the Index captures genuine workforce structure and plausible adoption patterns. However, validation anchors reflect current contexts—career transitions from human-only labor markets and technology-sector AI usage—which may not fully capture adoption dynamics in non-technology sectors. See Appendix A for validation methodology and anchor constraints.

\subsection{Measurement Scope}
The Index measures exposure from production AI systems deployed in enterprise contexts—automation platforms, copilots, and industry-specific tools—rather than frontier model capabilities tested on academic benchmarks. This focuses on immediate business impact: tools reshaping work today.
Recent evaluations like GDPval~\cite{openai2025gdpval} and APEX~\cite{vidgen2025apex} test whether frontier models can perform isolated professional tasks at human-equivalent quality. These benchmarks measure task-level capability; our framework measures systemic workforce exposure across 151 million workers. Both dimensions matter for workforce planning. A model may perform well on benchmarks yet not reshape work until packaged into deployable tools, integrated into workflows, and adopted at scale. By cataloging 13,000+ production systems, the Iceberg Index captures where AI is positioned to affect work now, providing conservative estimates of immediate exposure. Future work could examine alignment between high-exposure occupations and benchmark performance. See Appendix A for tool cataloging methodology.

\subsection{Modeling Choices}
Key assumptions affecting interpretation:
\begin{itemize}
\item \textbf{Wage-value weighting:} Enables consistent comparison across occupations but smooths over job quality dimensions (stability, autonomy, career progression) and masks within-occupation heterogeneity where some tasks face high automation potential while others remain human-intensive.
\item \textbf{Skill transferability:} The Index assumes skills demonstrated in one occupational context transfer to others, establishing an upper bound on exposure. This enables scenario analysis where transferability can be varied, but may overestimate near-term exposure if domain-specific adaptation proves more challenging than assumed.
\item \textbf{Digital AI focus:} Analysis covers cognitive and administrative automation where deployment is observable. Physical robotics excluded due to immature adoption data; see Appendix C for scope rationale.
\end{itemize}
See Appendices A and C for detailed data sources, implementation choices, and computational infrastructure.

\subsection{Causal Interpretation}
Validation is correlational rather than causal. Agreement with observed adoption patterns increases confidence but does not prove exposure drives outcomes. External factors—state investment, infrastructure, regulation—mediate how capability translates to impact. This limitation is inherent to forward-looking analysis: policymakers cannot wait for causal evidence of disruption before preparing responses. The Index provides the best available prospective indicator, validated against current patterns.

\subsection{Complementary Role}
The Index complements traditional labor metrics (GDP, unemployment, wages) rather than replacing them. Traditional metrics track realized outcomes; the Iceberg Index reveals potential exposure before adoption crystallizes. Together they enable comprehensive workforce planning combining retrospective assessment and prospective intelligence.

\section{Conclusion}
The Iceberg Index measures where AI technical capabilities overlap with human occupational skills across 151 million workers, providing forward-looking intelligence to complement traditional workforce metrics that track employment outcomes after disruption occurs.

The analysis reveals a substantial measurement gap. Current AI adoption concentrates in technology occupations representing 2.2\% of labor market wage value. Yet AI technical capability extends to cognitive and administrative tasks spanning 11.7\% of the labor market—approximately \$1.2 trillion in wage value across finance, healthcare, and professional services. This fivefold exposure difference is geographically distributed nationwide rather than concentrated in coastal hubs, indicating that workforce preparation strategies based on visible technology-sector signals may substantially undercount transformation potential. Validation against independent data confirms the framework captures plausible workforce patterns.

The Index measures technical exposure—where AI can perform occupational tasks—not displacement outcomes. Actual workforce impacts depend on firm strategies, worker adaptation, and policy choices. By simulating how capabilities might spread under different scenarios, Project Iceberg enables states to test interventions before committing resources, transforming workforce planning from reactive crisis management to strategic foresight. Future work will model adoption dynamics, extend to physical automation as robotics mature, and integrate task-level quality benchmarks.

The Iceberg Index provides measurable intelligence for critical workforce decisions: where to invest in training, which skills to prioritize, how to balance infrastructure with human capital. It reveals not only visible disruption in technology sectors but the larger transformation beneath the surface. By measuring exposure before adoption reshapes work, the Index enables states to prepare rather than react—turning AI into a navigable transition.

\section{Acknowledgement}
We thank Oak Ridge National Lab compute award for the simulation infrastructure; Anthropic for providing API credits; and MIT MLC for funding Ayush Chopra.

\bibliographystyle{abbrvnat}
\bibliography{main}  

\clearpage
\appendix
\section*{Appendix A: Data Sources and Framework}
The Iceberg Index combines four key components: skills required by occupations, automatability of those skills by AI systems, likelihood of adoption across industries, and economic value of affected work. Our framework integrates multiple data sources to model each component.

\paragraph{Skills Required by Occupations}: The O*NET Database~\cite{onet2024} provides both quantitative and qualitative skill requirements across 923 occupations. Numerical ratings cover Work Activities (41 behaviors like "Analyzing Data," "Interacting with Computers"), Skills (35 capacities like "Programming," "Critical Thinking"), and Knowledge areas (33 domains like "Economics and Accounting," "Engineering and Technology") with importance ratings (1-5 scale) and level ratings (0-7 scale) based on expert analyst assessments and incumbent worker surveys. Rich textual descriptions capture daily work activities, work values, cognitive/psychomotor/technical requirements, and detailed task descriptions that reveal what workers actually do beyond numerical scores. Categorical classifications include work context factors, education requirements, and physical demands. This multi-modal approach captures both measurable skill requirements and qualitative aspects of work that together define what each occupation entails.

\paragraph{AI System Capabilities:} We catalog over 13,000 production-ready AI tools from Model Context Protocol implementations (software development tools), the Zapier automation platform (workflow systems), and the OpenTools directory (specialized applications). These represent AI capabilities that can be packaged into deployable systems for specific occupational contexts, rather than raw frontier model performance on academic benchmarks. To align these tools with the Bureau of Labor Statistics (BLS) skill taxonomy, we develop a semi-automated mapping pipeline. Specifically, we use in-context learning with large language models to infer which skills each tool can perform, based on its task descriptions and metadata. For calibration, we split 600 occupations as training prompts and reserve 300 occupations for validation, enabling the model to learn skill-task correspondences from human-labeled BLS tasks. The model then predicts skill coverage for the tool set, which is manually reviewed to ensure consistency. This hybrid approach allows us to generate skill capability profiles for AI tools at scale, while retaining human oversight to correct systematic errors and validate uncertain cases. The result is a skill-level capability matrix that enables direct comparison between human job requirements and AI system functionality across the same dimensions.

\paragraph{Economic Value of Work}: The Bureau of Labor Statistics Occupational Employment and Wage Statistics~\cite{bls2024} provides employment counts and median wages for all 923 occupations across states and metropolitan areas, covering approximately 134 million wage and salary workers. The American Community Survey~\cite{acs2025} contributes geographic and demographic distributions for 151 million workers across 3,000 counties, including age, education, industry context, and commuting patterns. This enables calculation of total wage value potentially affected by automation and supports aggregation from individual workers to state and national levels.

\section*{Appendix B: Measuring Industry Concentration of AI Exposure}

While the Iceberg Index captures the overall level of AI-exposed work in a state, it does not reveal whether that exposure is concentrated in a few dominant industries or distributed broadly across the economy. This structural distinction has significant policy implications: states with concentrated exposure can focus on sector-specific strategies, while those with distributed exposure must prepare for system-wide transitions that span multiple sectors.

To quantify this structure, we apply a measure of industry concentration widely used in labor economics: the Herfindahl–Hirschman Index (HHI). Originally developed to assess market concentration, the HHI has been validated in labor-market research as a reliable tool for measuring employer dominance and sectoral risk~\cite{Azar2020LaborMarketConcentration, RevisitedHHI2023}. Here, we adapt it to capture how AI-related disruption is distributed across industries within each state.

The method proceeds as follows. For each state, we calculate the Iceberg Index contribution from each industry, reflecting the share of total exposed wage value attributable to that sector. We then express each industry’s contribution as a share of the state's total exposed wages, square these shares, and sum them to obtain a single concentration score. Formally, let $E_i$ be the exposed wage value for industry $i$, and let $E_{\text{total}} = \sum_i E_i$ denote the total exposed wage value in the state. We define the share of exposure from industry $i$ as $s_i = E_i / E_{\text{total}}$, and compute the HHI as:

\[
\text{HHI} = \sum_{i=1}^{N} s_i^2
\]

where $N$ is the number of industries. The HHI ranges from near zero (perfect diversification) to 1 (total concentration in one industry). For interpretability, we report values on a 0–10,000 scale, as is standard in economic literature. In our analysis, state-level HHI scores typically fall between 1400 and 1900.

To aid interpretation, we group states into three categories based on their HHI scores. Values below 1580 indicate distributed exposure across many sectors; values between 1580 and 1737 suggest moderately concentrated risk; and values above 1738 reflect states where most exposure is channeled into a few dominant industries. These thresholds loosely mirror antitrust classifications and align with the empirical distribution of concentration in our dataset.

This approach reveals clear regional patterns. States in the Northeast Corridor tend to show tightly clustered exposure, often concentrated in finance and technology. In contrast, states in the Manufacturing Belt exhibit more distributed profiles, with risk spread across production, logistics, administration, and related services. Smaller states vary widely: Iowa, for example, has one of the most diversified exposure profiles in the country (HHI = 1463), while Delaware concentrates a large share of its exposure in finance (HHI = 1741), despite having a comparable Iceberg Index.

Using a validated measure like the HHI ensures that structural comparisons across states are both rigorous and interpretable. It provides policymakers with a deeper view of AI disruption—not just how much exposure exists, but how that exposure is likely to manifest and propagate within each state’s economy.

\section*{Appendix C: Implementation Details}
The computational framework integrates two components to enable population-scale analysis.

\paragraph{(a) Large Population Models}: Implemented through the AgentTorch platform~\cite{chopra2024flame, chopra2025lpm}, providing the simulation engine for population-scale agent-based modeling. The platform supports efficient simulation of 151 million heterogeneous agents with distinct skill profiles while capturing interaction dynamics across geographic and occupational networks. LPMs have been validated through successful applications in epidemic response~\cite{chopra2023epidemiology}, biosecurity analysis~\cite{adiga2024biosecurity}, and supply chain resilience modeling~\cite{romero2021vaccine}, demonstrating their capability to handle complex population-scale dynamics. The framework incorporates behavioral adaptation mechanisms and technology adoption dynamics for realistic workforce response modeling. Core capabilities include scenario analysis through forward simulation, parameter estimation using real-world workforce data, and policy evaluation across multiple variables.

\paragraph{(b) High-Performance Computing Infrastructure}: Deployment on Oak Ridge National Laboratory's Frontier supercomputer~\cite{frontier2022}, one of the world's fastest computing systems, provides the computational execution layer. The infrastructure delivers massively parallel processing across thousands of compute nodes with GPU acceleration that supports large-scale optimization and analysis operations. This computational capacity transforms previously intractable population modeling challenges into feasible real-time policy analysis tools. These integrated components provide the computational foundation for the Iceberg Index and its policy scenario testing functionality, enabling dynamic analysis of workforce transformation scenarios under different policy conditions.



\end{document}